\documentclass[11pt]{article}
\usepackage[dvips]{graphicx}
\usepackage{psfrag}
\usepackage{amssymb}
\usepackage{amsmath}
\usepackage{amscd}
\usepackage{latexsym}

\catcode`@=11
\makeatletter\renewcommand{\section}{\@startsection
{section}{1}{\z@}{-3.5ex plus -1ex minus
    -.2ex}{2.3ex plus .2ex}{\large\bf }}

\makeatletter\renewcommand{\subsection}{\@startsection{subsection}{2}{\z@}{-3.25ex
plus -1ex minus
   -.2ex}{1.5ex plus .2ex}{\bf }}

\numberwithin{equation}{section}
\newcounter{saveeqn}

\catcode`@=12

\def\a{\alpha}
\def\b{\beta}
\def\ga{\gamma}
\def\Ga{\Gamma}
\def\de{\delta}

\def\ve{\varepsilon}

\def\om{\omega}
\newcommand{\C}{\mathbb C}
\newcommand{\R}{\mathbb R}
\newcommand{\Hbb}{\mathbb H}

\newcommand{\Acal}{{\cal A}}
\newcommand{\Ecal}{{\cal E}}

\newcommand{\Fcal}{{\cal F}}

\newcommand{\Ccal}{{\cal C}}
\newcommand{\X}{{\cal X}}
\newcommand{\Y}{{\cal Y}}
\newcommand{\Z}{{\cal Z}}
\newcommand{\gfrak}{{\mathfrak g}}
\newcommand{\mfrak}{{\mathfrak m}}
\newcommand{\hfrak}{{\mathfrak h}}
\newcommand{\su}{{\mathfrak{su}}}
\newcommand{\so}{{\mathfrak{so}}}
\newcommand{\fsp}{{\mathfrak{sp}}}
\newcommand{\fu}{{\mathfrak{u}}}
\newcommand{\spin}{{\mathfrak{spin}}}

\newcommand{\mh}{{\widehat{\smash{\mu}}}}
\newcommand{\nh}{{\widehat{\smash{\nu}}}}
\newcommand{\Gh}{{\widehat{\Gamma}}}
\newcommand{\Rh}{{\widehat{R}}}
\newcommand{\Ih}{{\hat{I}}}
\newcommand{\lrc}{\mathop{\lrcorner}}
\newcommand{\hra}{\mathop{\hookrightarrow}}
\def\1{{\bar 1}}
\def\im{\textrm{i}}
\def\diff{\textrm{d}}
\def\pa{\mbox{$\partial$}}
\def\sfrac#1#2{{\textstyle\frac{#1}{#2}}}
\def\+{\dagger}
\def\={\ =\ }
\def\und{\qquad\textrm{and}\qquad}
\def\and{\quad\textrm{and}\quad}
\def\with{\quad\textrm{with}\quad}
\def\for{\quad\textrm{for}\quad}

\begin{document}

\begin{titlepage}
\setcounter{page}{0}

\hspace{2.0cm}

\begin{center}

{\Large\bf
Instantons on Special Holonomy Manifolds
}

\vspace{12mm}

{\Large Tatiana~A.~Ivanova and Alexander~D.~Popov}
\\[15mm]

\noindent {\em
Bogoliubov Laboratory of Theoretical Physics, JINR\\
141980 Dubna, Moscow Region, Russia}\\
{Email: ita, popov@theor.jinr.ru}

\vspace{20mm}

\begin{abstract}
\noindent We consider cones over manifolds admitting real Killing spinors and instanton
equations on connections on vector bundles over these manifolds. Such cones are manifolds
with special (reduced) holonomy. We generalize the scalar ansatz for a connection
proposed by Harland and N\"olle~\cite{HN} in such a way that instantons are parameterized
by constrained matrix-valued functions. Our ansatz reduces instanton equations to a
matrix model equations which can be further reduced to Newtonian mechanics with particle
trajectories obeying first-order gradient flow equations. Generalizations  to
K\"ahler-Einstein manifolds and resolved Calabi-Yau cones are briefly discussed. Our
construction allows one to associate quiver gauge theories with special holonomy
manifolds via dimensional reduction.
\end{abstract}

\end{center}
\end{titlepage}

\section{Introduction}

Instantons in four dimensions~\cite{1} are nonperturbative Bogomolny-Prasad-Sommerfeld
(BPS) configurations solving first-order anti-self-duality equations for gauge fields
which imply the full Yang-Mills equations. They are important objects in modern field
theory~\cite{2,3}. Generalization of Yang-Mills instantons to higher dimensions, proposed
in~\cite{4} and studied in~\cite{5}-\cite{11} (for more literature see references
therein),
 is important both in mathematics~\cite{9,10} and string theory~\cite{12,13}. Some of their
solutions on spaces $\R^n$ with $n{>}5$ were obtained in~\cite{5,14,15}. Constructions of
solutions to the instanton equations on more general curved homogeneous manifolds as well
as on cylinders and cones over them were considered in~\cite{16,17}. The construction on
coset spaces, many of which admit Killing spinors~\cite{18}, was generalized to cones
over manifolds with real Killing spinors~\cite{HN}, not necessarily homogeneous (see
also~\cite{19} about instantons on Calabi-Yau cones and their resolutions). All these
solutions were lifted to solutions of heterotic supergravity equations supplemented by
the Bianchi identity~\cite{20,21,19,HN,22}.

Riemannian manifolds ($M,g^{}_M$) with real Killing spinors\footnote{A Killing spinor on a Riemannian
manifold $N$ is a spinor field $\epsilon$ which satisfies $\nabla_L\epsilon= \im\lambda L\cdot\epsilon$
for all tangent vectors $L$, where $\nabla$ is the spinor covariant derivative, $\cdot$ is Clifford
multiplication and $\lambda$ is a constant. If $\lambda = 0$ then the spinor is called parallel, and
$N$ is a manifold with special (reduced) holonomy.} often occur in string theory compactifications
(see e.g.~\cite{HN,20,21,22} and references therein). These manifolds were classified in~\cite{18}.
Besides the round spheres they are
\begin{itemize}
 \item nearly K\"ahler 6-manifolds $M$, SU(3)-structure
\item nearly parallel 7-manifolds $M$, $G_2$-structure
\item Sasaki-Einstein $(2m+1)$-manifolds $M$, SU($m$)-structure
\item 3-Sasakian $(4m+3)$-manifolds $M$, Sp($m$)-structure
\end{itemize}
All these manifolds have a connection with a non-vanishing torsion and admit a non-integrable $H$-structure
mentioned above, i.e. a reduction of the structure group SO($n$) of the tangent bundle $TM$ to $H\subset \ $SO($n$).
The above manifolds are equipped with canonical 3-form $P$ and 4-form $Q$ defined via the Killing spinors.

Recall that instanton equations on an $(n+1)$-dimensional Riemannian manifold $X$ can be introduced as follows.
Suppose there exists a 4-form $Q$ on $X$. Then there exists an $(n-3)$-form $\ast Q$, where $\ast$
is the Hodge operator on $X$. Let $\Acal$ be a connection on a bundle over $X$ with the curvature $\Fcal$.
Then the generalized anti-self-duality equation on the gauge field $\Fcal$ is~\cite{9,10}
\begin{equation}\label{1.1}
\ast\Fcal + \ast Q\wedge\Fcal =0\ .
\end{equation}
{}For $n+1>4$ these equations can be defined on manifolds $X$ with {\it special
holonomy}, i.e. such that the holonomy group $G$ of the Levi-Civita connection on the
tangent bundle $TX$ is a subgroup in the group SO($n+1$). On such manifolds any solution
of eq.(\ref{1.1}) satisfies to the Yang-Mills equation. The instanton equation
(\ref{1.1}) is also well-defined on manifolds $X$ with non-integrable $G$-structures but
then (\ref{1.1}) implies the Yang-Mills equation with torsion. This torsion term vanishes
on manifolds with real Killing spinors~\cite{HN}.

In this paper, we mostly consider $X=\Ccal (M)$, where $M$ is a manifold with real Killing spinors
 and $\Ccal (M)$ is a cone over $M$ with the metric
\begin{equation}\label{1.2}
 g^{}_X=\diff r^2+r^2g_M=e^{2\tau}(\diff\tau^2 + g_M) \for r:=e^{\tau}\ .
\end{equation}
{}From (\ref{1.2}) it follows that the cone $\Ccal (M)$ is conformally equivalent to the cylinder
\begin{equation}\label{1.3}
 Z=\R\times M
\end{equation}
with the metric
\begin{equation}\label{1.4}
 g^{}_Z=\diff\tau^2 +g_M\ .
\end{equation}
{}Furthermore, one can show~\cite{HN} that the equation (\ref{1.1}) on the cone $X=\Ccal
(M)$ is related with instanton equation on the cylinder $Z=\R\times M$ as follows
\begin{equation}\label{1.5}
 \ast_X\Fcal + \ast_X Q_X\wedge\Fcal = e^{(n-3)\tau}(\ast_Z\Fcal + \ast_Z Q_Z\wedge\Fcal )=0\ ,
\end{equation}
where $n+1=\,$dim$\,\Ccal (M)=\,$dim$\,Z$. In other words, eq.(\ref{1.1}) on $\Ccal (M)$
is equivalent to the equation on $\R\times M$ after rescaling (\ref{1.2}) of the metric.
That is why in the following we will consider the instanton equation
\begin{equation}\label{1.6}
\ast\Fcal + \ast Q_Z\wedge\Fcal = 0
\end{equation}
on the cylinder $Z$ over $M$. Here we omit the index $Z$ in the star operator. Note that
components of $\Fcal$ on the cone can be obtained from those on the cylinder simply via
rescaling (\ref{1.2}).

In this paper, we generalize the results~\cite{HN} of Harland and N\"olle on
investigating instantons on cones over manifolds with Killing spinors. First, in section
2, we collect various facts concerning nearly K\"ahler, nearly parallel $G_2$,
Sasaki-Einstein and 3-Sasakian manifolds $M$ following mainly the description
in~\cite{HN}. We describe metrics on $M$, canonical connections and various $q$-forms
($q=1,2,...)$ as well as their extension to the cylinder $Z=\R\times M$. Then, in section
3, we introduce an ansatz for a gauge potential $\Acal$ which reduces the instanton
equation (\ref{1.6}) on $\R\times M$ to a matrix equations on $\R$. Resolution of natural
algebraic constraints on the matrices yields further reduction to a set of first-order
equations on functions depending on $\tau\in\R$. These equations are gradient flow
equations describing BPS-type trajectories in Newtonian mechanics of particles moving in
$\R^N$, where $N$ is the number of functions parameterizing matrices in the ansatz for a
gauge potential $\Acal$. Solutions to these equations give instanton solutions of the
Yang-Mills equations on $\R\times M$ and their extension to the cone $\Ccal (M)$.
Finally, in section 4, we discuss some generalizations of our construction allowing one
to associate quiver gauge theories with such special holonomy manifolds as
K\"ahler-Einstein manifolds and resolved Calabi-Yau cones.

\section{Manifolds with Killing spinors}

\subsection{Nearly K\"ahler 6-manifolds}

Consider the cylinder (\ref{1.3}) with the metric (\ref{1.4}), where $M$ is a nearly
K\"ahler 6-manifold. It is defined as a manifold with a 2-form $\om$ and a 3-form $P$
such that
\begin{equation}\label{2.1}
\diff\om = 3\ast_MP\and \diff P=2\om\wedge\om =:4Q\ .
\end{equation}
For a local orthonormal co-frame $\{e^a\}$ on $M$ one can choose
\begin{equation}\label{2.2}
\om = e^{12}+e^{34}+e^{56}\and P=e^{135}+e^{164}-e^{236}-e^{245}\ ,
\end{equation}
where $a=1,...,6,\ e^{a_1...a_l}:=e^1\wedge...\wedge e^l$, and get
\begin{equation}\label{2.3}
\ast_MP = e^{145}+e^{235}+e^{136}-e^{246}\ ,\quad Q=e^{1234}+e^{1256}+e^{3456}\ .
\end{equation}
Here $\ast_M$ denotes the Hodge operator on $M$. On $Z$ one can introduce the
4-form
\begin{equation}\label{2.4}
Q_Z=\diff\tau\wedge P + Q\ ,
\end{equation}
which is used in the instanton equation (\ref{1.6}).

The canonical connection $\tilde\Gamma$ on $M$, which is a metric-compatible connection
with totally antisymmetric (intrinsic) torsion, has SU(3) structure group. This
connection has components
\begin{equation}\label{2.5}
\tilde\Gamma^c_{ab}=\Gamma^c_{ab}+\sfrac12P_{cab}\ ,
\end{equation}
where $\Gamma^c_{ab}$ are components of the Levi-Civita connection and
\begin{equation}\label{2.6}
 P_{abc}= T^a_{bc}
\end{equation}
are components of the torsion $T^a=\sfrac12\, T^a_{bc}e^b\wedge e^c$ defined from the Cartan
structure equations
\begin{equation}\label{2.7}
 \diff e^a + \tilde\Gamma^a_b\wedge e^b = T^a
\end{equation}
for basis 1-forms $e^a$.

Note that the structure group of $M$ is SU(3) (or its subgroup) and $P$ induces a $G_2$-structure
on $Z$ since SU(3)$\subset G_2$. Recall that $\gfrak_2 = \su(3)\oplus\mfrak$, dim$\,\mfrak =6$, and
one can define the generators of $G_2$ as 7$\times$7 matrices from $\so(7)$ with the commutation relations
\begin{equation}\label{2.8}
[I_i, I_j]\=f^k_{ij}\, I_k \ ,\qquad [I_i, I_a]\=f^b_{ia}\, I_b\und
[I_a, I_b]\=f^i_{ab}\, I_i + f^c_{ab}\, I_c\ ,
\end{equation}
where $I_i\in \su(3)$, $I_a\in\mfrak$ and $f$'s are structure constants of $\gfrak_2$. One
can choose~\cite{HN}
\begin{equation}\label{2.9}
\begin{aligned}
 I_{ia}^b =& f_{ia}^b\ ,  & I_{ia}^0=&-I_{i0}^a = 0\ ,\\
 I_{ab}^c =& \sfrac{1}{2}f^c_{ab}\ , & I_{a0}^b=& -I_{ab}^0 = \delta_a^b\ ,
\end{aligned}
\end{equation}
and obtain
\begin{equation}\label{2.10}
 P_{abc}=-f^c_{ab}\ .
\end{equation}
Introducing $\mu =(0,a)$, we can denote matrices in (\ref{2.9}) as $I^{\mu}_{i\nu}$ and $I^{\mu}_{a\nu}$.
The extension of the canonical connection $\tilde\Gamma$ to $Z$ has the same components (\ref{2.5}) and
can be written as
\begin{equation}\label{2.11}
 \tilde\Gamma = \tilde\Gamma^iI_i
\end{equation}
with $I_i$ given in (\ref{2.9}).

\subsection{Nearly parallel $G_2$-manifolds}

Let us consider the cylinder (\ref{1.3}) over a nearly parallel $G_2$-manifold $M$. It is
defined as a manifold
with a 3-form $P$ (a $G_2$-structure) preserved by the group $G_2\subset\,$SO(7) such that
\begin{equation}\label{2.12}
 \diff P =\gamma\, \ast_MP
\end{equation}
for some constant $\gamma\in\R$. For a local orthonormal co-frame $e^a$, $a=1,...,7$, on
$M$ one can choose
\begin{equation}\label{2.13}
P = e^{123} + e^{145} - e^{167} + e^{246} + e^{257} + e^{347} - e^{356}
\end{equation}
and therefore
\begin{equation}\label{2.14}
\ast_M P=: Q = e^{4567} +e^{2367} -e^{2345} +e^{1357} +e^{1346} +e^{1256} -e^{1247}.
\end{equation}
It is easy to see that for the choice (\ref{2.13}) one obtains $\diff P=4Q$, i.e. $\gamma =4$.
The 4-form $Q_Z$ on $Z$ can be chosen similar to (\ref{2.4}) as
\begin{equation}\label{2.15}
 Q_Z=\diff\tau\wedge P + Q\ .
\end{equation}
This form defines a Spin(7)-structure on $Z$.

One can define generators of the group Spin(7) via the structure constants $f_{ij}^k,
f_{ia}^b$ and $f_{ab}^c$ of the group Spin(7) by using the decomposition
$\spin(7)=\gfrak_2\oplus\mfrak$,  as 8$\times$8 matrices
$I_i=(I_{i\nu}^{\mu})\in \gfrak_2$ and $I_a=(I_{a\nu}^{\mu})\in \mfrak\ ,
\ $dim$\,\mfrak =7$, $\mu =(0,a)$.
The generators $I_i, I_a$ have the same form as in (\ref{2.9}) but with structure
constants $f$'s of Spin(7).

The canonical connection $\tilde\Gamma$ on $M$ is not changed after its extension to $Z$ and has the components
\begin{equation}\label{2.16}
  \tilde\Gamma = \tilde\Gamma^iI_i\quad\Rightarrow\quad
\tilde\Ga^i_aI_{ib}^c=\tilde\Gamma^c_{ab} = \Gamma^c_{ab}+\sfrac13\, P_{abc}
\end{equation}
with torsion components
\begin{equation}\label{2.17}
 T^a_{bc}=\sfrac23\, P_{abc}\ .
\end{equation}

\subsection{Sasaki-Einstein manifolds}

Consider now the cylinder (\ref{1.3}) with the metric (\ref{1.4}), where $M$ is a
Sasaki-Einstein manifold. It is a $(2m+1)$-dimensional manifold such that the cone $\Ccal
(M)$ with the metric (\ref{1.2}) is a Calabi-Yau $(m+1)$-fold~\cite{BG}. Such manifolds
$M$ have the structure group SU$(m)\subset\,$SO$(2m+1)$ and the holonomy group of $\Ccal
(M)$ is SU$(m+1)$. Sasaki-Einstein manifolds are endowed with 1-, 2-, 3- and 4-forms
$\eta , \om , P$ and $Q$, which can be defined in an orthonormal basis $e^1, e^a,
a=2,...,2m+1$, as
\begin{equation}\label{2.18}
\eta=e^1\ ,\quad \omega = e^{23}+e^{45}+\dots +e^{2m\,2m+1}\ ,
\quad P = \eta\wedge\omega\and Q = \frac12\omega\wedge\omega \ .
\end{equation}
One can check that $\eta\lrcorner\,\omega=0$ and
\begin{equation}\label{2.19}
\diff\eta=2\om \ ,\quad\diff\ast_M\om =2m\ast_M\eta\ ,\quad\diff P=4Q\and \diff \ast_M Q
= (2m-2)\ast_M P\ .
\end{equation}

The metric on $Z$ has the form (\ref{1.4}) with
\begin{equation}\label{2.20}
 g_M=(e^1)^2+\exp(2h)\delta_{ab}e^ae^b\ .
\end{equation}
Note that for the value of $h$ such that
\begin{equation}\label{2.21}
\exp(2h)=\frac{2m}{m+1}\ ,
\end{equation}
the torsion of the canonical connection on $M$ (and on $Z$) becomes
antisymmetric~\cite{HN}, but we keep the one-parameter family (\ref{2.20}) of Sasakian
metric including the case $h=0$ when the metric is Einstein. Components of the canonical
connection $\tilde\Gamma$ are
\begin{equation}\label{2.22}
 \tilde\Ga_{\mu a}^b=\Ga_{\mu a}^b + \frac{1}{m}P_{\mu ab}\ ,\quad -\tilde\Ga_{\mu a}^1=
\tilde\Gamma_{\mu 1}^a =\Gamma_{\mu 1}^a+ P_{\mu 1a}\ ,
\end{equation}
where $\mu =(1,a)$ and the torsion of $\tilde\Ga$ is
\begin{equation}\label{2.23}
 T^1= P_{1\mu\nu}e^\mu\wedge e^\nu \and
T^a = \frac{m+1}{2m} P_{a\mu\nu}e^\mu\wedge e^\nu \ .
\end{equation}
As 4-form $Q_Z$ on $Z$ one can take~\cite{HN}
\begin{equation}\label{2.24}
 Q_Z=\exp(2h)\diff\tau\wedge P + \exp(4h)Q\ ,
\end{equation}
where $P$ and $Q$ are given in (\ref{2.18}).

Let $\hat\mu =(0,\mu )=(0,1,a)$. Then one can define generators of the group
$\su(m+1)=\su(m)\oplus\mfrak$ as $(2m+1)\times (2m+1)$ antisymmetric matrices $I_i=(I^{\
\mh}_{i\ \nh})\in \su(m)$ and $I_{\mu}=(I^{\ \mh}_{\mu\ \nh})\in\mfrak$ such that
non-vanishing components are
$$
I^b_{ia}=f^b_{ia}\ ,
$$
\begin{equation}\label{2.25}
I^b_{1a}=-\frac{1}{m}\, P_{1ab}=(m+1)f^b_{1a}\ ,\quad -I^0_{ab}=I^b_{a0}=\de^b_a\ ,
\end{equation}
$$
I^1_{ab}=-I^b_{a1}=-P_{1ab}=\sfrac12\, f^1_{ab}\ ,
$$
where $f^b_{ia}$, $f^1_{ab}$ and $f^b_{1a}$ are parts of the structure constants of $\su(m+1)$. In
terms of these matrices the canonical connection $\tilde\Ga$ on $M$ pulled-back to $Z$ can be written as
\begin{equation}\label{2.26}
\tilde\Ga =\tilde\Ga^i I_i=e^\mu \tilde\Ga^i_\mu I_i\ .
\end{equation}

\subsection{3-Sasakian manifolds}

Let us now consider the cylinder (\ref{1.3}) over a 3-Sasakian manifold $M$. It is
defined as a $(4m+3)$-dimensional manifold such that the cone $\Ccal (M)$ over it is a
hyper-K\"ahler $(4m+4)$-manifold~\cite{BG}, i.e. the holonomy group of $\Ccal (M)$ is
Sp$(m+1)$. The structure group of $M$ is Sp($m$) and any 3-Sasakian manifold can be
endowed with three 1-forms $\eta^\alpha$, three 2-forms $\omega^\alpha$, a 3-form $P$ and
a 4-form $Q$, $\alpha=1,2,3$~\cite{BG}.  In a local orthonormal co-frame $e^\alpha,e^a$,
$a=4,\dots,4m+3$, these forms can be written as
\begin{equation}\label{2.27}
 \begin{aligned}
  \eta^1&= e^1\ ,\quad \omega^1 = e^{45}+e^{67}+\dots +e^{4m\,4m+1}+e^{4m+2\,4m+3}\ , \\
  \eta^2&= e^2\ ,\quad \omega^2 = e^{46}-e^{57}+\dots +e^{4m\,4m+2}-e^{4m+1\,4m+3}\ , \\
  \eta^3&= e^3\ ,\quad \omega^3 = e^{47}+e^{56}+\dots +e^{4m\,4m+3}+e^{4m+1\,4m+2}\ , \\
P&=\sfrac13\eta^{123}+\sfrac13\eta^\a\wedge\om^\a\and Q=\sfrac16\, \om^\a\wedge\om^\a\ .
 \end{aligned}
\end{equation}
The forms $\eta^\a$ and $\om^\a$ satisfy the equations
\begin{equation}\label{2.28}
  \diff\eta^\alpha = \varepsilon_{\alpha\beta\gamma}\eta^\beta\wedge\eta^\gamma + 2\omega^\alpha \ , \quad
  \diff\omega^\alpha = 2\varepsilon_{\alpha\beta\gamma}\eta^\beta\wedge\omega^\gamma\ .
\end{equation}

We introduce indices $\mu = (\alpha , a)$ and $\mh =(0,\mu )=(0,\a , a)$. Using the splitting
\begin{equation}\label{2.29}
 \fsp(m+1)=\fsp(m)\oplus\mfrak\ ,\quad \mbox{dim}\,\mfrak=4m+3\ ,
\end{equation}
one can introduce generators  $I_i=(I^{\ \mh}_{i\ \nh})\in \fsp(m)$ and $I_{a}=(I^{\
\mh}_{a\ \nh})\in\mfrak$ of the group Sp($m+1$) as matrices from $\so(4m+4)$. One can take
them so that non-vanishing components are~\cite{HN}
$$
I^b_{ia}=f^b_{ia}\ ,
$$
\begin{equation}\label{2.30}
I^{\gamma}_{\a\b}=-\varepsilon_{\alpha\beta\gamma}=-3P_{\alpha\beta\gamma}=\sfrac12\,f^{\gamma}_{\a\b}\ ,
\quad I^{\b}_{\a 0}=\delta^\b_\a\ ,
\end{equation}
$$
I^\a_{ab}=-\om^\b_{ab}=-3P_{\a ab}=\sfrac12\, f^\a_{ab}\ ,\quad I^b_{a0}=\de^b_a\ ,
$$
where $f$'s are the structure constants of the group Sp($m+1$).

Note that the metric on $Z=\R\times M$ has the form (\ref{1.4}) with a one-parameter
family
\begin{equation}\label{2.31}
 g_M=\de_{\a\b}e^\a e^\b + \exp(2h)\, \de_{ab}e^a e^b
\end{equation}
of metrics on $M$.
The 4-form $Q_Z$ can be chosen as
\begin{equation}\label{2.32}
 Q_Z=\sfrac16(\exp(4h)\om^\a\wedge\om^\a + \exp(2h)\ve_{\a\b\ga}\om^\a\wedge\eta^\b\wedge\eta^{\ga} +
2\,\exp(2h)\diff\tau\wedge\eta^\a\wedge\om^\a + 6\diff\tau\wedge\eta^{123})\ .
\end{equation}
In terms of the matrices (\ref{2.30}) the canonical connection $\tilde\Gamma$ on the cylinder $Z$ over a
3-Sasakian manifold $M$ can be written as
\begin{equation}\label{2.33}
 \tilde\Ga = \tilde\Ga^iI_i=e^{\mu}\tilde\Ga^i_\mu I_i\ .
\end{equation}
It is related with the Levi-Civita connection $\Ga$ by formulae
\begin{equation}\label{2.34}
 -\tilde\Ga^{\nu}_{\mu\a} = \tilde\Ga^{\a}_{\mu\nu}=\Ga^{\a}_{\mu\nu} + 3P_{\a\mu\nu}\ ,
\quad\tilde\Ga^{b}_{\mu a}=\Ga^{b}_{\mu a}\ ,
\end{equation}
and has the torsion
\begin{equation}\label{2.35}
 T^\a = 3P_{\a\mu\nu}e^{\mu\nu}\and T^a = \sfrac32 P_{a\mu\nu}e^{\mu\nu}
\end{equation}
which is antisymmetric for the choice $\exp (2h)=2$ in the metric.

\section{Instantons in higher dimensions}

\subsection{Reduction to matrix equations}

Recall that for all cases of manifolds $M$ considered in section 2 the instanton equation on the cone $\Ccal (M)$ is
equivalent to the equation
\begin{equation}\label{3.1}
 \ast\Fcal + \ast Q_Z\wedge\Fcal =0
\end{equation}
on the cylinder $Z=\R\times M$ with the metric $g_Z=\diff\tau^2 + g_M$. The explicit form of the 4-form $Q_Z$ on $Z$ was
written down for all cases in section 2. Let us denote by $G$ the holonomy group\footnote{This holonomy group $G$ is the
group $G_2$, Spin(7), SU($m{+}1$) and Sp($m{+}1$) for cones over nearly K\"ahler, nearly parallel $G_2$, Sasaki-Einstein
and 3-Sasakian manifolds $M$, respectively.} of the Levi-Civita connection on $\Ccal (M)$ and by $H$ the structure
group\footnote{This  structure
group is the group SU(3), $G_2$, SU($m$) and Sp($m$) for nearly K\"ahler, nearly parallel
$G_2$, Sasaki-Einstein and 3-Sasakian manifolds, respectively.} of the canonical
connection $\tilde\Ga$ on $M$ (and also on $Z$). For the Lie algebras $\gfrak
=\,$Lie$\,G$ and $\hfrak =$\,Lie$\,H$ we have
\begin{equation}\label{3.2}
 \gfrak =\hfrak \oplus\mfrak\ ,
\end{equation}
where $\mfrak$ is an orthogonal complement of $\hfrak$ in $\gfrak$. Let $e^0=\diff\tau$ and $e^\mu$ be an orthonormal
basis of $T^*Z$. Then $e^\mu$ form a basis of $T^*M\subset T^*Z$ and their linear span can be identified with the
vector space $\mfrak$.

Consider the generators
\begin{equation}\label{3.3}
[\Ih_i, \Ih_j]\=f^k_{ij}\, \Ih_k \ ,\qquad [\Ih_i, \Ih_\mu]\=f^\nu_{i\mu}\, \Ih_\nu\and
[\Ih_\mu, \Ih_\nu]\=f^i_{\mu\nu}\, \Ih_i + f^\sigma_{\mu\nu}\, \Ih_\sigma
\end{equation}
acting on the space $V$ of an irreducible representation of $G$. These generators satisfy
the same commutation relations as the generators $I_i, I_a$. In section 2 we wrote down
the realization of these generators via the embedding $\gfrak\subset \so(n+1)$ with
$n+1=\,$dim$Z$ as acting (via infinitesimal rotations) on tangent spaces of $Z$. For this
special representation we omit hats and note that $\tilde\Ga =\tilde\Ga^iI_i$ is the
canonical connection whose curvature
\begin{equation}\label{3.4}
 \tilde R = \diff\tilde\Ga + \tilde\Ga\wedge\tilde\Ga = (\diff\tilde\Ga^i + \sfrac12\,
 f^i_{jk}\tilde\Ga^j\wedge\tilde\Ga^k)I_i
\end{equation}
satisfies the instanton equation (\ref{3.1}) (see~\cite{HN}).

Note that instead of tangent bundle $TZ$ one can consider an arbitrary vector bundle
${\cal V}\to Z$ with the structure group $G$ such that fibres are representations $V$
(real, complex or quaternionic) of the group $G$. For simplicity, we consider irreducible
representations $V$ of the group $G$. The bundle ${\cal V}\to Z$ is associated with the
principal bundle $P(Z,G)$. Since $H$ is a closed subgroup of $G$, it also acts on fibres
of $\cal V$, but in general after restriction to $H\subset G$ the representation $V$
decomposes into a sum of irreducible representations $V_{q_r}$ of $H$ such that
$V=\oplus_rV_{q_r}$. We denoted the generators of the group $G$ in the representation $V$
as $\hat I_i$, $\hat I_\mu$, where $\hat I_i\in \hfrak$ and $\hat I_\mu\in \mfrak$ for
the splitting (\ref{3.2}).  Consider a connection
\begin{equation}\label{3.5}
 \Gh := \tilde\Ga^i\hat I_i
\end{equation}
on the bundle ${\cal V}$. In general, it is a reducible connection. Here $\tilde\Ga^i$
are components of the canonical connection on the tangent bundle $TZ$. It is obvious that
the curvature $\Rh = \diff\Gh + \Gh\wedge\Gh$ of $\Gh$ also satisfies the instanton
equation (\ref{3.1}).

Let us consider matrix-valued functions $X_\mu (\tau)\in\ $End($V$) and introduce a
connection
\begin{equation}\label{3.6}
 \Acal :=\Gh + X_\mu e^\mu
\end{equation}
on the vector bundle ${\cal V}\to Z$. Note that for matrices $X_\mu$ depending on all
coordinates of $Z$, (\ref{3.6}) is a general form of a connection on the bundle ${\cal
V}\to Z$. For $X_\mu$ depending only on $\tau$, the instanton equation (\ref{3.1}) will
be reduced to ordinary differential equations on matrices $X_\mu$.

Recall that
\begin{equation}\label{3.7}
\diff e^\mu = - \tilde\Ga^\mu_\nu\wedge e^\nu + T^\mu = -\tilde\Ga^i\wedge e^\nu f^\mu_{i\nu}+\sfrac12\,
T^\mu_{\sigma\nu}e^\sigma\wedge e^\nu \ ,
\end{equation}
where $f^\mu_{i\nu}$ are structure constants from (\ref{3.3}). From (\ref{3.6}) and (\ref{3.7})
it follows that
\begin{equation}\label{3.8}
\Fcal =
\diff\Acal{+}\Acal{\wedge}\Acal = \diff\Gh{+}\Gh\wedge\Gh{+}\sfrac12([X_\mu , X_\nu]{+}T^\sigma_{\mu\nu}X_\sigma )e^\mu{\wedge}e^\nu{+}
\dot X_\nu e^0{\wedge}e^\nu{+}\tilde\Ga^i{\wedge}e^\mu ([\Ih_i , X_\mu]{-} f^\nu_{i\mu} X_\nu ),
\end{equation}
where $\dot X_\nu =\frac{\diff X_\nu}{\diff\tau}$. Note that $\Rh = \diff\Gh + \Gh\wedge\Gh$ satisfies to eq.(\ref{3.1})
and $\Fcal$ solves the instanton equation (\ref{3.1}) on $Z$ if there are satisfied the following matrix equations:
\begin{equation}\label{3.9}
[\Ih_i, X_\mu ]= f^\nu_{i\mu}X_\nu\ ,
\end{equation}
\begin{equation}\label{3.10}
[X_\mu , X_\nu ] + T^\sigma_{\mu\nu}X_\sigma = N^\sigma_{\mu\nu}\dot X_\sigma +
f^i_{\mu\nu}N_i(\tau )\ .
\end{equation}
Here $N^\sigma_{\mu\nu}$ are some constants which we shall specify below for each case,
$N_i$ are some $v(\hfrak )$-valued functions defined by eqs.(\ref{3.10}) after resolving
the algebraic constraint equations (\ref{3.9}) and substituting their solutions $X_\mu$
into (\ref{3.10}). Here $v: \gfrak\to\ $End($V$) is a representation of $\gfrak$. For
$X_\mu$ satisfying (\ref{3.9})-(\ref{3.10}), we have
\begin{equation}\label{3.8a}
\Fcal=\diff\Gh +\Gh\wedge\Gh + \sfrac12\,N_if^i_{\mu\nu}e^{\mu\nu} +
\dot X_\sigma (e^{0\sigma} + \sfrac12\,N^\sigma_{\mu\nu}e^{\mu\nu} )\ ,
\end{equation}
where the term with $f^i_{\mu\nu}$ also satisfies (\ref{3.1}) due to properties of
$f^i_{\mu\nu}$ and the last term with $\dot X_\sigma$ satisfies (\ref{3.1}) for choices
of $N^\sigma_{\mu\nu}$ specified below for each considered case. Note that the constraint
equations (\ref{3.9}) for some examples of groups $G$ and $H$ were discussed and resolved
e.g. in~\cite{23,24} in the context of the equivariant dimensional reductions on coset
spaces $G/H$. For special cases of the ansatz (\ref{3.6}) instanton solutions were
obtained e.g. in~\cite{16, 17, HN}. Peculiar property of such $\tau$-dependent solutions
is that they can be lifted to gauge 5-brane solutions of heterotic supergravity equations
as was shown e.g in~\cite{19, HN, 22}.

\subsection{Reduction for nearly K\"ahler and nearly parallel $G_2$ manifolds}

Consider a manifold $M$ which is nearly K\"ahler (dim$\,M=n=6$) or nearly parallel $G_2$
(dim$\,M=n=7$). For both cases $\mu = a =1,...,n$ with $n=6$ or $n=7$. Note that the
2-forms
\begin{equation}\label{3.11}
e^{0a}-\frac{1}{2\rho}\,P_{abc}e^{bc}
\end{equation}
solve the instanton equation (\ref{3.1}) on $Z=\R\times M$ for $Q_Z$ and $P$ given in
section 2. Here $\rho =2$ for $n=6$ and  $\rho =3$ for $n=7$. The generators $\Ih_a$
introduced in (\ref{3.3}) are images of the 2-forms  (\ref{3.11}) under the
metric-induced  isomorphism $\Lambda^2 Z\cong \so(7)\supset\gfrak_2\supset\mfrak$ for
$n=6$ and $\Lambda^2 Z\cong \so(8)\supset \spin(7) \supset\mfrak$ for $n=7$.

For both the nearly K\"ahler and nearly parallel $G_2$ cases we have
\begin{equation}\label{3.12}
T^a_{bc} = - f^a_{bc}\and N^a_{bc}=\sfrac12\,f^a_{bc}\ ,
\end{equation}
where $N^a_{bc}$ are defined by comparing the components of $\Fcal$ in (\ref{3.8a}) and
the explicit form (\ref{3.11}) of (parts of) anti-self-dual forms on $Z$. Thus, we obtain
the following matrix equations:
\begin{equation}\label{3.13}
[\Ih_i, X_a ]= f^b_{ia}X_b\ ,
\end{equation}
\begin{equation}\label{3.14}
[X_a , X_b ] = f_{ab}^c(X_c+\sfrac12\dot X_c) + f_{ab}^iN_i(\tau)\ .
\end{equation}
Substituting the ansatz $X_a=\phi \Ih_a$ with a real function $\phi (\tau)$, we see that
(\ref{3.13}) are satisfied and (\ref{3.14}) are reduced to the equation
\begin{equation}\label{3.15}
\dot\phi = 2\phi (\phi -1)
\end{equation}
obtained in~\cite{HN} and for $N_i$ we obtain $N_i = \phi^2 \Ih_i$. More general
equations can be obtained by choosing more general solution of the constraint equations
(\ref{3.13}). Such solutions for different choice of groups $G$ and $H$ were discussed
e.g. in~\cite{23, 24}. Constructing solutions to eqs. (\ref{3.9}), (\ref{3.10}) goes
beyond the scope of this short article. This task will be considered elsewhere.

\subsection{Reduction for Sasaki-Einstein manifolds}

Consider $Z=\R\times M$ with a Sasaki-Einstein manifold $M$. In this case $\mu =(1,a)$ with
$a=2,...,2m+1$ and the 2-forms
\begin{equation}\label{3.16}
 e^{01}-\frac{1}{m+1}\,\om_{ab}e^{ab}\and\exp(h)( e^{0a}+\om_{ab} e^{1b})
\end{equation}
solve the instanton equations (\ref{3.1}) with $Q_Z$ given in (\ref{2.24}). From (\ref{2.23}),
(\ref{3.8}) and (\ref{3.16}) we obtain
\begin{equation}\label{3.17}
T^1_{ab}=-f^1_{ab}\ ,\quad T^a_{1b}=-f^a_{1b}\ ,\quad N^1_{ab}=\frac{1}{m+1}\,f^1_{ab}\and
N^a_{1b}=\frac{m}{m+1}\,f^a_{1b}\ .
\end{equation}
Substituting (\ref{3.17}) into (\ref{3.9})-(\ref{3.10}), we obtain
\begin{equation}\label{3.18}
[\Ih_i, X_1 ]=0\ ,\quad [\Ih_i, X_a ]= f^b_{ia}X_b\ ,
\end{equation}
\begin{equation}\label{3.19}
[X_1 , X_a ] = f_{1a}^b(X_b+\frac{m}{m+1}\,\dot X_b) \ .
\end{equation}
\begin{equation}\label{3.20}
[X_a , X_b ] = f_{ab}^1 (X_1 + \frac{1}{m+1}\,\dot X_1)+f_{ab}^iN_i(\tau)\ .
\end{equation}
If we choose $X_1=\chi \Ih_1$ and $X_a=\psi \Ih_a$ then (\ref{3.18}) are satisfied, for $N_i$
we obtain $N_i=\psi^2\Ih_i$ and (\ref{3.19})-(\ref{3.20}) are reduced to the equations
\begin{equation}\label{3.21}
\dot\chi = (m+1)(\psi^2 -\chi )\and\dot\psi =\frac{m+1}{m}\psi (\chi -1)
\end{equation}
coinciding with those obtained in~\cite{HN}. Note that (\ref{3.21}) is a gradient flow equation
$$\dot x^i=g^{ij}\frac{\pa W}{\pa x^j} \for W=(x^1)^2-2x^1(x^2)^2+2(x^2)^2-1,\ i,j=1,2,
$$
for the metric on $\R^2$ of the form
$$\diff s^2=g_{ij}\diff x^i\diff x^j=\frac{2}{m+1}(\diff x^1)^2 + \frac{4m}{m+1}(\diff x^2)^2$$
with $x^1:=\chi$ and $x^2:=\psi$~\cite{HN}.

\subsection{Reduction for 3-Sasakian manifolds}

For $Z=\R\times M$  with a 3-Sasakian manifold $M$ we have $\mu =(\a ,a), \a =1,2,3$ and $a=4,...,4m+3$.
One can check that the image of the generators $\Ih_a$ from (\ref{3.3}) under the map into $\Lambda^2Z\cong
\so(4m+4)\supset \fsp(m+1)\supset\mfrak$ is given by the 2-forms
\begin{equation}\label{3.22}
 e^{0\a}-\sfrac13\,\ve_{\a\b\ga}e^{\b\ga}\and\exp(h)( e^{0a}+\om_{ab}^\a e^{\a b})
\end{equation}
which satisfy to eq.(\ref{3.1}) with $Q_Z$ given in (\ref{2.32}).
From (\ref{2.30}), (\ref{2.35}), (\ref{3.8}) and (\ref{3.22}) it follows that
\begin{equation}\label{3.23}
T^\a_{ab}=-\sfrac32\,f^\a_{ab}\ ,\quad T^b_{a\b}=f^b_{a\b}\ ,\quad T^\a_{\b\ga}=-f^\a_{\b\ga}\ ,
\end{equation}
\begin{equation}\label{3.24}
N^b_{a\b}=f^b_{a\b}\and
N^\a_{\b\ga}=\sfrac12\,f^\a_{\b\ga}
\end{equation}
with other components vanishing. Substituting (\ref{3.23})-(\ref{3.24}) into (\ref{3.9})-(\ref{3.10}), we obtain
\begin{equation}\label{3.25}
[\Ih_i, X_\a ]=0\ ,\quad [\Ih_i, X_a ]= f^b_{ia}X_b\ ,
\end{equation}
\begin{equation}\label{3.26}
[X_\a , X_\b ] = f_{\a\b}^\ga(X_\ga +\sfrac12\,\dot X_\ga) \ , \quad [X_a , X_\b ] = f_{a\b}^b(X_b +\dot X_b)\ ,
\end{equation}
\begin{equation}\label{3.27}
[X_a , X_b ] = f_{ab}^\a X_\a + f_{ab}^iN_i(\tau)\ .
\end{equation}

If we choose the ansatz $X_\a = \chi\Ih_\a$ and $X_a = \psi\Ih_a$ then (\ref{3.25}) will be satisfied
identically, from (\ref{3.27}) we obtain $N_i=\psi^2\Ih_i$ and (\ref{3.26})-(\ref{3.27}) reduce to the equations
\begin{equation}\label{3.28}
\dot\chi = 2\chi(\chi -1)\ ,\quad \dot\psi =\psi (\chi -1)\and\chi = \psi^2\ ,
\end{equation}
where the last algebraic equation follows from (\ref{3.27}). These equations coincide
with those obtained in~\cite{HN}. Thus, our ansatz (\ref{3.6}) which leads to matrix
equations (\ref{3.9})-(\ref{3.10}) generalizes the ``scalar'' ansatz of the
paper~\cite{HN} and allows one to obtain more general instanton solutions. However,
obtaining explicit instanton solutions  lies beyond the scope of our paper.

\section{Generalizations: quiver bundles}

\subsection{Smaller groups $H$}

Recall that we considered nearly K\"ahler, nearly parallel $G_2$, Sasaki-Einstein and
3-Sasakian manifolds $M$ with the structure groups SU(3), $G_2$, SU($m$) and Sp($m$),
respectively, following to Harland and N\"olle who considered in their ansatz~\cite{HN}
exactly the above groups with generators in the defining vector representation of the
group SO($n+1$)$\ \supset H$, i.e. $V=\R^{n+1}$ with $n=6, 7, 2m+1$ and $4m+3, m=1,
2,...\ $. However, the group $H$ can be smaller than the above-mentioned Lie groups, i.e.
often $H$ lies {\it inside} the group SU(3), $G_2$, SU($m$) and Sp($m$), respectively. In
this case, the constraint equations (\ref{3.9}) become weaker and allow more degrees of
freedom in matrices $X_a$. For instance, for the nearly K\"ahler coset space
\begin{equation}\label{4.1}
 M=\mbox{SU}(3)/\mbox{U}(1)\times\mbox{U}(1)
\end{equation}
the structure group is $H=\ $U(1)$\times$U(1) that increase the number of functions
parameterizing the ansatz (\ref{3.6}) even for vector representation
$V=\R^7\cong\R\oplus\C^3$ with $\gfrak_2 \supset
\su(3)=\fu(1){\oplus}\fu(1){\oplus}\mfrak$. Writing the ansatz (\ref{3.6}) in terms of
$\su (3)$-valued matrices $X_a$, one can resolve (\ref{3.9}) as
\begin{equation}\label{4.2}
\begin{aligned}
X_1&\={\small\begin{pmatrix}
0&\ 0&\!\!-\phi_1\\0&\ 0&0\\\bar\phi_1&\ 0&0\end{pmatrix}} \ ,\quad
X_3\={\small\begin{pmatrix}
0&\!\!-\bar\phi_2&\ 0\\\phi_2&0&\ 0\\0&0&\ 0\end{pmatrix}} \ ,\quad
X_5\={\small\begin{pmatrix}
0&\ 0&0\\0&\ 0&\!\!-\bar\phi_3\\0&\ \phi_3&0\end{pmatrix}} \ , \\[8pt]
X_2&\={\small\begin{pmatrix}
0&0&\im\phi_1\\0&0&0\\\im\bar\phi_1&0&0\end{pmatrix}} \ ,\quad
X_4\=-{\small\begin{pmatrix}
0&\im\bar\phi_2&\ 0\\\im\phi_2&0&\ 0\\0&0&\ 0\end{pmatrix}} \ ,\quad
X_6\=-{\small\begin{pmatrix}
0\ &0&0\\0\ &0&\im\bar\phi_3\\0\ &\im\phi_3&0\end{pmatrix}} \ ,
\end{aligned}
\end{equation}
where $\phi_1, \phi_2, \phi_3$ are complex-valued functions of $\tau$ and
the generators $I_{7,8}$ of the subgroup U(1)${\times}$U(1) of
SU(3) are chosen in the form
\begin{equation}\label{4.3}
I_7\=-\im\,
{\small\begin{pmatrix}0&\ 0&0\\0&\ 1&0\\0&\ 0&\!\!-1\end{pmatrix}}
\und
I_8\=\frac{\im}{\sqrt{3}}\,
{\small\begin{pmatrix}2&0&0\\0&\!\!-1&0\\0&0&\!\!-1\end{pmatrix}}\ .
\end{equation}
Substituting (\ref{4.2}) into (\ref{3.14}), we obtain equations
\begin{equation}\label{4.4}
 \dot\phi_1=-2\phi_1 + 2\bar\phi_2\bar\phi_3\ ,\quad
\dot\phi_2=-2\phi_2 + 2\bar\phi_1\bar\phi_3\ ,\quad
\dot\phi_3=-2\phi_3 + 2\bar\phi_1\bar\phi_2
\end{equation}
and constraints
\begin{equation}\label{4.5}
 N_7=\Phi\,I_7\ ,\quad N_8=-\sqrt{3}\Phi\,I_8 \with \Phi =\phi_1\bar\phi_1=\phi_2\bar\phi_2=\phi_3\bar\phi_3
\end{equation}
for a proper normalization of the structure constants. From  (\ref{4.5}) we see that complex-valued functions
$\phi_1, \phi_2, \phi_3$ can differ only in their phase parts. For $\phi_1 = \phi_2 = \phi_3 =:\phi$ eqs.(\ref{4.4})
reduce to eq.(\ref{3.15}) on a real-valued function $\phi$.

\subsection{Reducible representations of $H$ and quiver bundles}

Similar situation takes place for nearly parallel $G_2$-manifolds, where as an example
one can consider the Aloff-Wallach space SU(3)/U(1) with the structure group $H=\ $U(1)
(see the second paper in~\cite{17} for discussion of solving eqs.(\ref{3.13})), and also
for Sasaki-Einstein and 3-Sasakian manifolds  the structure group $H$ can be a closed
subgroup of SU($m$) and Sp($m$), respectively. Even more freedom appears if one considers
an irreducible representation $V$ of the holonomy group $G$ of the cone $\Ccal (M)$ which
decomposes into a sum of irreducible representations $V_{q_r}$ of the group $H$,
\begin{equation}\label{4.6}
 V=\mathop{\oplus}^\ell_{r=1}V_{q_r}\with\mathop{\sum}^\ell_{r=1}q_r=q
\end{equation}
so that
\begin{equation}\label{4.7}
\Ih_i=
\begin{pmatrix}
I^{q_1}_i&0&\dots&0\\
0&\ddots&&\vdots\\
\vdots&&\ddots&0\\
0&\dots&0&I^{q_\ell}_i
\end{pmatrix}\ .
\end{equation}
Here $I^{q_r}_i$ are generators of $q_r\times q_r$ irreducible representations $V_{q_r}$
of $H$ and $V\cong\C^q$ (or $\R^q$ or $\Hbb^q\cong\R^{4q}$). If we assume that $H$
contains a maximal abelian subgroup of $G$ then the remaining generators $\Ih_a$ of $G$
in this representation have the off-diagonal form\footnote{If $H$ does not contain a
maximal abelian subgroup of $G$ or there is a subgroup in $G$ commuting with $H$ then
$\Ih_a$ in (\ref{4.8}) will contain diagonal terms $I_a^{q_{rr}}$ with $r=1,...,\ell$.}
\begin{equation}\label{4.8}
\Ih_a=
\begin{pmatrix}0&I^{q_{12}}_a&\dots&I^{q_{1\ell}}_a \\
I^{q_{21}}_a&0&\ddots&\vdots\\
\vdots&\ddots&\ddots&I^{q_{\ell-1\,\ell}}_a \\
I^{q_{\ell 1}}_a&\dots&I^{q_{\ell\,\ell -1}}_a&0
\end{pmatrix}\ ,
\end{equation}
where $I_a^{q_{rs}}$ are $q_r\times q_s$ matrices (cf.~\cite{11}).

Thus, one can associate a bounded quiver\footnote{A quiver $Q=(Q^0,Q^1)$ is an oriented
graph, i.e. a set of vertices $Q^0$ with a set $Q^1$ of arrows between the vertices (see
e.g.~\cite{25}). A path in $Q$ is a sequence of arrows in $Q^1$ which compose. A relation
of the quiver is a formal finite sum of paths. In our case vertices correspond to vector
bundles ${\cal V}_{q_r}$ with fibres $V_{q_r}$ and arrows correspond to morphisms ${\cal
V}_{q_s}\to {\cal V}_{q_r}$ of vector bundles.} satisfying a set of relations $R$ to the
ansatz (\ref{3.6}) for a connection on a vector bundle ${\cal V}\to\Ccal (M)$ over nearly
K\"ahler, nearly parallel $G_2$, Sasaki-Einstein and 3-Sasakian manifolds. In the
simplest case of generators (\ref{4.7}) the matrices $X_a$ solving the constraint
equations are obtained from (\ref{4.8}) by substituting $\phi_{{rs}}I^{q_{rs}}_a$ instead
of  $I^{q_{rs}}_a$, where  $\phi_{{rs}}$ are complex functions of $\tau$. Important fact
is that the space $\Ccal (M)$ is not homogeneous and therefore quivers and quiver bundles
can appear in dimensional reduction without $G$-equivariance condition studied earlier
e.g. in~\cite{23,24}. Recall that another way in which quiver gauge theories arise as
low-energy effective field theories in string theory is through considering cones and
orbifolds with conical singularities and placing D-branes at the orbifold
singularities~\cite{26}. Our constructions can be lifted as in~\cite{22} to heterotic
strings and provide a description of NS5-branes and gauge NS5-branes. It would be of
interest to study further this brane interpretation and its possible relations with
constructions of~\cite{26}.

\subsection{K\"ahler-Einstein manifolds and quiver gauge theories}

As another example related with quiver gauge theories we consider the manifold
\begin{equation}\label{4.9}
 \Y =\Sigma\times \X \ ,
\end{equation}
 where $\Sigma$ and $\X$ are 2-dimensional and $2k$-dimensional K\"ahler-Einstein manifolds with the K\"ahler
form $\om$ on $\Sigma$ and $\Omega$ on $\X$. Let $\Gh$ be the canonical $\fu (k)$-valued
Levi-Civita connection on $\X$,
\begin{equation}\label{4.10}
 \Gh=\Ga^i\Ih_i\with \Ih_i\in \fu (k)\ .
\end{equation}
We consider U($k$) as a closed subgroup of the Lie group SU($k+1$). Let $V\cong\C^q$ be
an irreducible representation of the group SU($k+1$) decomposed into a sum of irreducible
representations $V_{q_r}\cong\C^{q_r}$ of the group  U($k$) as in (\ref{4.7}) and ${\cal
V}\to \X$ is a holomorphic vector bundle over $\X$ associated with the bundle $P(\X ,$
U($k$)) of hermitian frames on $\X$. This bundle has the connection (\ref{4.10}) which is
reducible according to (\ref{4.7}), ${\cal V}=\oplus_r {\cal V}_{q_r}$.

Consider now $\ell$ complex vector bundles $E_1,...,E_\ell$ over $\Sigma$ with unitary
connections $A^1,...,A^\ell$ and ranks $N_1,...,N_\ell$. Introduce a complex vector
bundle $\Ecal =\oplus_r\, E_r{\otimes}{\cal V}_{q_r}$ over $\Sigma\times \X$ of rank
\begin{equation}\label{4.11}
 N=\mathop{\sum}^\ell_{r=1}N_rq_r
\end{equation}
and assume $c_1(\Ecal )=0$ without loss of generality, so that the structure group of $\Ecal$ is SU($N$).
The matrices
\begin{equation}\label{4.12}
\tilde I_i:=
\begin{pmatrix}{\bf 1}_{N_1}\otimes I^{q_1}_i &0& \dots &0\\
0&\ddots&&\vdots\\
\vdots&&\ddots&0\\
0&\dots&0&{\bf 1}_{N_\ell}\otimes I^{q_\ell}_i
\end{pmatrix}
\end{equation}
are generators of a reducible unitary representation of the
group U($k$) on the complex vector space $\widetilde V\cong \C^N$.
Introduce a gauge connection
\begin{equation}\label{4.13}
A:=
\begin{pmatrix}A^1\otimes{\bf 1}_{q_1}&0&\dots &0\\
0&\ddots&&\vdots\\
\vdots&&\ddots&0\\
0&\dots &0&A^\ell\otimes{\bf 1}_{q_\ell}
\end{pmatrix}
\end{equation}
on the bundle $E:=\oplus_r\, E_r\otimes \C^{q_r}$ over $\Sigma$. It is obvious from
(\ref{4.13}) that $[A , \tilde I_i]=0$.

On the bundle $\Ecal\to \Y$ we introduce a connection
\begin{equation}\label{4.14}
\Acal =A+\Ga^i\tilde I_i+X_a e^a\ ,
\end{equation}
where $X_a\in\su (N)$ are matrices which depend only on coordinates of $\Sigma$ and $e^a$
is the basis of 1-forms on $\X$, $a=1,...,2k$. Note that
\begin{equation}\label{4.15}
 \diff\,e^a=-\Ga^a_b\wedge e^b = - f^a_{ib}\Ga^i\wedge e^b\ .
\end{equation}
Using  (\ref{4.15}), for the curvature $\Fcal$ of the connection  (\ref{4.14}) we obtain
\begin{equation}\label{4.16}
 \Fcal = \diff\Acal +\Acal\wedge\Acal = F + \widetilde R + (\diff X_a+ [A, X_a] )\wedge e^a +\sfrac{1}{2}\,
 [X_a, X_b]\, e^a\wedge e^b + ([I_i, X_a]-f^b_{ia}\, X_b ) \, \Ga^i\wedge e^a \ ,
\end{equation}
where
\begin{equation}\label{4.17}
 F = \diff A +A\wedge A\ ,\quad \widetilde R=\diff\widetilde\Ga  + \widetilde\Ga\wedge \widetilde\Ga \and
 \widetilde\Ga:=\Ga^i\tilde I_i\ .
\end{equation}
Suppose that $X_a$ satisfy the constraints
\begin{equation}\label{4.18}
[ \tilde I_i , X_a]=f^b_{ia} X_b
\end{equation}
and impose on $\Fcal$ the Hermitian-Yang-Mills equations~\cite{6}
\begin{equation}\label{4.19}
 \Fcal^{0,2}=0\quad\Rightarrow\quad \bar\pa X_a+[A^{0,1},X_a]=0\ ,\quad [Y_{\bar A}, Y_{\bar B}]=0\ ,
\end{equation}
\begin{equation}\label{4.20}
 (\om +\Omega )\lrc\Fcal =0\quad\Rightarrow\quad\om^{\a\b}\Fcal_{\a\b} + \lambda\tilde I_0 + \Omega^{ab}[X_a, X_b]=0\ .
\end{equation}
Here $\bar\pa + A^{0,1}$ is the anti-holomorphic part of the covariant derivative on
$\Sigma$,
$$Y_\1 :=\sfrac12(X_1 +\im\,X_{k+1}),..., Y_{\bar k} :=\sfrac12(X_k +\im\,X_{2k})\ ,$$
the constant $\lambda$ is proportional to the scalar curvature of the K\"ahler-Einstein
manifold $\X , \om_{\a\b}$ and $\Omega_{ab}$ are components of the K\"ahler forms on
$\Sigma$ and $\X ,\ \tilde I_0$ is the $\fu (1)$ generator in the decomposition $\fu (k)
= \fu (1)\oplus \su (k)\subset\su (k+1)$, $\a ,\b =1,2$ and $A, B,... = 1,...,k$. We see
that  (\ref{4.19})-(\ref{4.20}) are the usual quiver vortex equations on $\Sigma$
(cf.~\cite{11, 24}).\footnote{Note that the last equations in (\ref{4.19}) correspond to
quiver relations.} For $k=1$ and $\X =\C P^1$, one can obtain~\cite{27} the standard
vortex equations on a Riemann surface $\Sigma$. One can generalize the above construction
by taking instead of $\Sigma$ a K\"ahler-Einstein manifold of dimension more than two.

\subsection{On instantons on smooth manifolds}

It is of interest to extend the ansatz for a connection $\Acal$ from cones to their
smooth resolutions as it was proposed in~\cite{19} as well as from  direct product
manifolds, such as $\Y$ in section 4.3, to irreducible smooth manifolds with warped
product metrics. This is possible.

{}For illustration we consider non-compact Calabi-Yau $(k+1)$-folds $\Z$ discussed
in~\cite{19}. They have a metric
\begin{equation}\label{4.21}
 \diff\tilde s^2=\frac{\diff r^2}{f^2(r)} + r^2\,f^2(r)\,\eta^2 +
2r^2\, \diff s^2_{KE}\ ,
\end{equation}
where
\begin{equation}\label{4.22}
 f^2 = 1- \Big(\frac{a^2}{r^2}\Big)^{k+1}\ ,
\end{equation}
$\diff s^2_{KE}$ is the standard K\"ahler metric on a  K\"ahler-Einstein manifold $\X$
which is the base manifold for a projection
\begin{equation}\label{4.23}
\pi : \X'\to \X
\end{equation}
from Sasaki--Einstein $(2k+1)$-manifold $\X'$ onto $\X$ and $\eta$ is the 1-form along
fibres of the projection (\ref{4.23}).

Note that
\begin{equation}\label{4.24}
 \diff\tilde s^2= r^2\, \diff s^2\ ,
\end{equation}
with
\begin{equation}\label{4.25}
 \diff s^2=\frac{\diff r^2}{r^2f^2} + f^2\,\eta^2 + 2\diff s^2_{KE}=\frac{\diff \tau^2}{f^2} + f^2\,\eta^2 + 2\diff s^2_{KE}\ ,
\end{equation}
i.e. $\diff\,\tilde s^2$ is conformally equivalent to $\diff s^2$. Singularity of the
transformation at $r{=}0$ is not essential since we are interested in Yang-Mills
instantons on the manifold with the metric  (\ref{4.21}) extendable smoothly at $r=0$.
Note also that the instanton equation on $\Z$ is invariant w.r.t. conformal
transformation,
\begin{equation}\label{4.26}
\tilde\ast F + \tilde\ast Q_Z\wedge F= \tilde\ast F + (\tilde\om
+\tilde\Omega)^{k-1}\wedge F = r^{2(k-1)}(\ast F + (\om +\Omega)^{k-1}\wedge F)=0\ ,
\end{equation}
since
\begin{equation}\label{4.27}
\tilde\om +\tilde\Omega =r^2(\om +\Omega)
\end{equation}
and $\tilde\ast = r^{2(k-1)}\ast$. Here $\om =\diff\tau\wedge\eta$ and $\Omega$ is the
K\"ahler form on $\X\hra\Z$.

Ansatz for $\Acal$ on the space $\Z'$ with the metric  (\ref{4.25}) is the same as in
(\ref{4.14}) and lead to the same reduction  (\ref{4.19})-(\ref{4.20}) of the instanton
equation. Solving these vortex equations, one obtains instantons on $\Z$. One can
simplify the task assuming that $\Acal_\eta$ and $X_a$ depend only on $\tau =\ln r$ and
choosing $\Acal_\tau =0$. Then  (\ref{4.19})-(\ref{4.20}) will be reduced to equations
similar to those which were considered in~\cite{19}.

\section{Conclusions}

We have examined in some detail the construction of instantons on cones $\Ccal (M)$ over
nearly K\"ahler and nearly parallel $G_2$-manifolds $M$ initiated in~\cite{16, 17} and
extended to cones over Sasaki-Einstein and 3-Sasakian manifolds in~\cite{HN, 19}. Having
at our disposal a reduced structure group $H$ of a manifold $M$ admitting real Killing
spinors and the holonomy group $G$ of the cone $\Ccal (M)$, we introduced a quiver bundle
$\cal V$ over $\Ccal (M)$, determined entirely by the representation theory of the group
$G$ and $H$, and introduced a proper connection $\Acal$ on this bundle. The ansatz for
$\Acal$ reduces the instanton equations on $\Ccal (M)$ to simpler matrix equations which
can be solved in many special cases. It is of interest to construct new instanton
solutions on $\Ccal (M)$ by using our generalized ansatz, and to lift them to solutions
of heterotic supergravity along the way considered in~\cite{HN, 22}.

We have also introduced a quiver bundle $\cal E$ over a K\"ahler-Einstein manifold of the
form $\Sigma\times \X$ and extended the Levi-Civita connection on the $2k$-dimensional
K\"ahler-Einstein manifold $\X$ to a connection on $\cal E$ parameterized by $2k$
matrices $X_a$. We established an equivalence between solutions of Hermitian-Yang-Mills
equations on $\Sigma\times \X$ and solutions of some quiver vortex equations on $\Sigma$.

Recall that regular Sasaki-Einstein manifolds are U(1)-bundles over K\"ahler-Einstein
manifolds and cones over them are Calabi-Yau spaces. Using this correspondence, we have
introduced a connection on a quiver bundle $\Ecal$ over smooth resolutions of
$(2k+2)$-dimensional Calabi-Yau cones. It is of interest to consider instantons on other
special holonomy manifolds.\footnote{Some instanton solutions on particular kinds of
$G_2$- and Spin(7)-manifolds were considered in~\cite{28}.} We hope to report on this in
the future.

\bigskip

\noindent
{\bf Acknowledgements}

\medskip

\noindent
 The authors thank the Institute for Theoretical Physics of Hannover University, where this work was
completed, for hospitality. This work was partially supported  by the Deutsche
Forschungsgemeinschaft, the Russian Foundation for Basic Research  and  the
Heisenberg-Landau program.


\newpage

\begin{thebibliography}{99}

\bibitem{HN}
  D.~Harland and C.~N\"olle,
  ``Instantons and Killing spinors,''
  JHEP {\bf 03} (2012) 082
  [arXiv:1109.3552 [hep-th]].
  %%CITATION = JHEPA,1203,082;%%

\bibitem{1}
  A.A.~Belavin, A.M.~Polyakov, A.S.~Schwartz and Yu.S.~Tyupkin,
  ``Pseudoparticle solutions of the Yang-Mills equations,''
  Phys.\ Lett.\  B {\bf 59} (1975) 85.
  %%CITATION = PHLTA,B59,85;%%

\bibitem{2}
  T.~Eguchi, P.B.~Gilkey and A.J.~Hanson,
  ``Gravitation, gauge theories and differential geometry,''
  Phys.\ Rept.\  {\bf 66} (1980) 213.
  %%CITATION = PRPLC,66,213;%%

\bibitem{3}
R.~Rajaraman, {\it Solitons and Instantons}, North-Holland, Amsterdam, 1984.

\bibitem{4}
  E.~Corrigan, C.~Devchand, D.B.~Fairlie and J.~Nuyts,
  ``First order equations for gauge fields in spaces of dimension
    greater than four,''
  Nucl.\ Phys.\ B {\bf 214} (1983) 452.
 %%CITATION = NUPHA,B214,452;%%

\bibitem{5}
  R.S.~Ward,
  ``Completely solvable gauge field equations in dimension
    greater than four,''\\
  Nucl.\ Phys.\ B {\bf 236} (1984) 381.
  %%CITATION = NUPHA,B236,381;%%

\bibitem{6}
  S.K.~Donaldson,
  ``Anti-self-dual Yang-Mills connections on a complex algebraic surface
    and stable vector bundles,''
  Proc.\ Lond.\ Math.\ Soc.\ {\bf 50} (1985) 1;

  S.K.~Donaldson,
  ``Infinite determinants, stable bundles and curvature,''\\
  Duke Math.\ J.\ {\bf 54} (1987) 231;

  K.K.~Uhlenbeck and S.-T.~Yau,
  ``On the existence of Hermitian-Yang-Mills connections on stable bundles
    over compact K\"ahler manifolds,''
  Commun.\ Pure Appl.\ Math.\ {\bf 39} (1986) 257.

\bibitem{7}
  M.~Mamone~Capria and S.M.~Salamon,
  ``Yang-Mills fields on quaternionic spaces,''\\
  Nonlinearity {\bf 1} (1988) 517;

  R.~Reyes~Carri\'on,
  ``A generalization of the notion of instanton,''
  Diff.\ Geom.\ Appl.\ {\bf 8} (1998) 1.
  %%CITATION = DGAPE,8,1;%%

\bibitem{8}
  L.~Baulieu, H.~Kanno and I.M.~Singer,
  ``Special quantum field theories in eight and other dimensions,''
  Commun.\ Math.\ Phys.\ {\bf 194} (1998) 149
  [arXiv:hep-th/9704167];
  %%CITATION = CMPHA,194,149;%%

M.~Blau and G.~Thompson,
  ``Euclidean SYM theories by time reduction and special holonomy manifolds,''
  Phys.\ Lett.\ B {\bf 415} (1997) 242
  [hep-th/9706225];
  %%CITATION = HEP-TH/9706225;%%

  B.S.~Acharya, J.M.~Figueroa-O'Farrill, B.J.~Spence and M.~O'Loughlin,
  ``Euclidean D-branes and higher-dimensional gauge theory,''
  Nucl.\ Phys.\  B {\bf 514} (1998) 583
  [arXiv:hep-th/9707118].
  %%CITATION = NUPHA,B514,583;%%


\bibitem{9}
  S.K.~Donaldson and R.P.~Thomas,
  ``Gauge theory in higher dimensions,''\\
  in: {\it The Geometric Universe},
  Oxford University Press, Oxford, 1998;

  S.K.~Donaldson and E.~Segal,
  ``Gauge theory in higher dimensions II'',\\
  arXiv:0902.3239 [math.DG].

\bibitem{10}
  G.~Tian,
  ``Gauge theory and calibrated geometry,''\\
  Ann.\ Math.\ {\bf 151} (2000) 193
  [arXiv:math/0010015 [math.DG]].
  %%CITATION = ANMAA,151,193;%%

\bibitem{11}
  A.D.~Popov,
  ``Non-Abelian vortices, super-Yang-Mills theory and Spin(7)-instantons,''\\
  Lett.\ Math.\ Phys.\ {\bf 92} (2010) 253
  [arXiv:0908.3055 [hep-th]];
  %%CITATION = ARXIV:0908.3055;%%

  A.D.~Popov and R.J.~Szabo,
  ``Double quiver gauge theory and nearly K\"ahler flux compactifications,''
JHEP {\bf 02} (2012) 033
  [arXiv:1009.3208 [hep-th]].
  %%CITATION = JHEPA,1202,033;%%

\bibitem{12}
  M.B.~Green, J.H.~Schwarz and E.~Witten,
  {\it Superstring theory},\\
  Cambridge University Press, Cambridge, 1987.

\bibitem{13}
K.~Becker, M.~Becker and J.H.~Schwarz, {\it String theory and M-theory: A modern introduction},
Cambridge University Press, Cambridge, 2007.

\bibitem{14}
  D.B.~Fairlie and J.~Nuyts,
  ``Spherically symmetric solutions of gauge theories in eight dimensions,''
  J.\ Phys.\ A {\bf 17} (1984) 2867;
  %%CITATION = JPAGB,A17,2867;%%

  S.~Fubini and H.~Nicolai,
  ``The octonionic instanton,''
  Phys.\ Lett.\ B {\bf 155} (1985) 369;
  %%CITATION = PHLTA,B155,369;%%

  A.D.~Popov,
  ``Anti-self-dual solutions of the Yang-Mills equations in $4n$-dimensions,''\\
  Mod.\ Phys.\ Lett.\  A {\bf 7} (1992) 2077;
  %%CITATION = MPLAE,A7,2077;%%

  T.A.~Ivanova and A.D.~Popov,
  ``(Anti)self-dual gauge fields in dimension $d{\ge}4$,''\\
  Theor.\ Math.\ Phys.\ {\bf 94} (1993) 225.
  %%CITATION = TMPHA,94,225;%%


\bibitem{15}
E.~Corrigan, P.~Goddard and A.~Kent,
  ``Some comments on the ADHM construction in $4k$-dimensions,''
  Commun.\ Math.\ Phys.\  {\bf 100} (1985) 1;
  %%CITATION = CMPHA,100,1;%%

E.K.~Loginov,
  ``Multi-instantons in higher dimensions and superstring solitons,''\\
  SIGMA {\bf 1} (2005) 002
  [arXiv:hep-th/0511262];
  %%CITATION = 00480,1,002;%%

 J.~Broedel, T.A.~Ivanova and O.~Lechtenfeld,
``Construction of noncommutative instantons in $4k$ dimensions,''
  Mod.\ Phys.\ Lett.\  A {\bf 23} (2008) 179
  [arXiv:hep-th/0703009].
  %%CITATION = MPLAE,A23,179;%%


\bibitem{16}
  T.A.~Ivanova, O.~Lechtenfeld, A.D.~Popov and T.~Rahn,
  ``Instantons and Yang-Mills flows on coset spaces,''
  Lett.\ Math.\ Phys.\  {\bf 89} (2009) 231
  [arXiv:0904.0654 [hep-th]];
 %%CITATION = LMPHD,89,231;%%

D.~Harland, T.A.~Ivanova, O.~Lechtenfeld and A.D.~Popov,\\
  ``Yang-Mills flows on nearly K\"ahler manifolds and $G_2$-instantons,''\\
  Commun. Math. Phys. {\bf 300} (2010) 185
  [arXiv:0909.2730 [hep-th]];
  %%CITATION = CMPHA,300,185;%%

D.~Harland and A.D.~Popov,
  ``Yang-Mills fields in flux compactifications on homogeneous manifolds
  with SU(4)-structure,''
    JHEP {\bf 02} (2012) 107
  [arXiv:1005.2837 [hep-th]].
 %%CITATION = ARXIV:1005.2837;%%


\bibitem{17}
I.~Bauer, T.A.~Ivanova, O.~Lechtenfeld and F.~Lubbe,
  ``Yang-Mills instantons and dyons on homogeneous $G_2$-manifolds,''
  JHEP {\bf 10} (2010) 044
  [arXiv:1006.2388 [hep-th]];
  %%CITATION = JHEPA,1010,044;%%

  A.S.~Haupt, T.A.~Ivanova, O.~Lechtenfeld and A.D.~Popov,\\
  ``Chern-Simons flows on Aloff-Wallach spaces and Spin(7)-instantons,''\\
  Phys.\ Rev.\ D \textbf{83} (2011) 105028,
  [arXiv:1104.5231 [hep-th]];
 %%CITATION = PHRVA,D83,105028;%%

K.-P.~Gemmer, O.~Lechtenfeld, C.~N\"olle and A.D.~Popov,
  ``Yang-Mills instantons on cones and sine-cones over nearly K\"ahler
  manifolds,''
  JHEP {\bf 09} (2011) 103
  [arXiv:1108.3951 [hep-th]].
  %%CITATION = JHEPA,1109,103;%%


\bibitem{18}
C.~B\"ar, ``Real Killing spinors and holonomy,''
Commun. Math. Phys. {\bf 154} (1993) 509.

\bibitem{19}
F.P.~Correia,
  ``Hermitian Yang-Mills instantons on Calabi-Yau cones,''\\
  JHEP {\bf 12} (2009) 004
  [arXiv:0910.1096 [hep-th]];
  %%CITATION = JHEPA,0912,004;%%

  F.P.~Correia,
  ``Hermitian Yang-Mills instantons on resolutions of Calabi-Yau cones,''\\
  JHEP {\bf 02} (2011) 054
  [arXiv:1009.0526 [hep-th]].
  %%CITATION = JHEPA,1102,054;%%


\bibitem{20}
  A.~Strominger,
  ``Heterotic solitons,''
  Nucl. Phys. B {\bf 343} (1990) 167;
  %[Erratum-ibid. B {\bf 353} (1991) 565];
  %%CITATION = NUPHA,B343,167;%%

  J.A.~Harvey and A.~Strominger,
  ``Octonionic superstring solitons,''\\
  Phys.\ Rev.\ Lett.\  {\bf 66} (1991) 549;
  %%CITATION = PRLTA,66,549;%%

T.A.~Ivanova,
  ``Octonions, self-duality and strings,''
  Phys.\ Lett.\  B {\bf 315} (1993) 277;
  %%CITATION = PHLTA,B315,277;%%

  M.~Gunaydin and H.~Nicolai,
   ``Seven-dimensional octonionic Yang-Mills instanton and its extension to an
  heterotic string soliton,''
  Phys. Lett.  B {\bf 351} (1995) 169
 %[Addendum-ibid.\  B {\bf 376}, 329 (1996)];
  [arXiv:hep-th/9502009].
  %%CITATION = PHLTA,B351,169;%%


\bibitem{21}
O.~Lechtenfeld, C.~N\"olle and A.D.~Popov,
  ``Heterotic compactifications on nearly K\"ahler manifolds,''
  JHEP {\bf 09} (2010) 074
  [arXiv:1007.0236 [hep-th]];
  %%CITATION = JHEPA,1009,074;%%

  C.~N\"olle,
  ``Homogeneous heterotic supergravity solutions with linear dilaton,''\\
  J.\ Phys.\ A  {\bf 45} (2012) 045402
  [arXiv:1011.2873 [hep-th]];
  %%CITATION = JPAGB,A45,045402;%%

   A.~Chatzistavrakidis, O.~Lechtenfeld and A.~D.~Popov,
  ``Nearly K\"ahler heterotic compactifications with fermion condensates,''
  arXiv:1202.1278 [hep-th].
  %%CITATION = ARXIV:1202.1278;%%

\bibitem{22}
K.-P.~Gemmer, A.S.~Haupt, O.~Lechtenfeld, C.~N\"olle and A.D.~Popov,
  ``Heterotic string plus five-brane systems with asymptotic AdS$_3$,''
  arXiv:1202.5046 [hep-th].
  %%CITATION = ARXIV:1202.5046;%%

\bibitem{BG}
 C.P.~Boyer and K.~Galicki,
  ``Sasakian geometry, holonomy, and supersymmetry,''\\
  arXiv:math/0703231.
  %%CITATION = MATH/0703231;%%

\bibitem{23}
O.~Garcia-Prada,
  ``Invariant connections and vortices,''\\
  Commun.\ Math.\ Phys.\  {\bf 156} (1993) 527;
  %%CITATION = CMPHA,156,527;%%

 L.~Alvarez-Consul and O.~Garcia-Prada,
  ``Dimensional reduction and quiver bundles,''\\
J. Reine Angew. Math. {\bf 556} (2003) 1
[arXiv:math/0112160].
  %%CITATION = MATH/0112160;%%

\bibitem{24}
A.D.~Popov and R.J.~Szabo,
  ``Quiver gauge theory of nonabelian vortices and noncommutative  instantons
  in higher dimensions,''
  J.\ Math.\ Phys.\  {\bf 47} (2006) 012306
  [arXiv:hep-th/0504025];
  %%CITATION = JMAPA,47,012306;%%

O.~Lechtenfeld, A.~D.~Popov and R.~J.~Szabo,
  ``Quiver gauge theory and noncommutative vortices,''
  Prog.\ Theor.\ Phys.\ Suppl.\  {\bf 171} (2007) 258
  [arXiv:0706.0979 [hep-th]];
  %%CITATION = PTPSA,171,258;%%

O.~Lechtenfeld, A.D.~Popov and R.J.~Szabo,
  ``SU(3)-equivariant quiver gauge theories and nonabelian vortices,''
  JHEP {\bf 08} (2008) 093
  [arXiv:0806.2791 [hep-th]];
  %%CITATION = JHEPA,0808,093;%%

B.P.~Dolan and R.J.~Szabo,
  ``Dimensional reduction and vacuum structure of quiver gauge theory,''
  JHEP {\bf 08} (2009) 038
  [arXiv:0905.4899 [hep-th]].
  %%CITATION = JHEPA,0908,038;%%

\bibitem{25}
M.~Auslander, I.~Reiten and S.O.~Smal\o,
{\it Representation theory of Artin algebras},
Cambridge University Press, Cambridge, 1995.

\bibitem{26}
M.R.~Douglas and G.W.~Moore,
``D-branes, quivers and ALE instantons,''
hep-th/9603167;
%%CITATION = HEP-TH 9603167;%%

C.V.~Johnson and R.C.~Myers,
``Aspects of Type~IIB theory on ALE spaces,''\\
Phys. Rev. D {\bf 55} (1997) 6382 [hep-th/9610140];
%%CITATION = HEP-TH 9610140;%%

M.R.~Douglas, B.~Fiol and C.~Romelsberger, ``The spectrum of BPS branes on a noncompact
Calabi-Yau,'' JHEP {\bf 09} (2005) 057 [hep-th/0003263].
%%CITATION = HEP-TH 0003263;%%

\bibitem{27}
  A.D.~Popov,
  ``Integrability of vortex equations on Riemann surfaces,''\\
  Nucl.\ Phys.\  B {\bf 821} (2009) 452
  [arXiv:0712.1756 [hep-th]];
  %%CITATION = NUPHA,B821,452;%%

A.D.~Popov,
  ``Non-abelian vortices on Riemann surfaces: an integrable case,''\\
  Lett.\ Math.\ Phys.\  {\bf 84} (2008) 139
  [arXiv:0801.0808 [hep-th]].
  %%CITATION = LMPHD,84,139;%%

\bibitem{28}
 S.~Miyagi,
``Yang-Mills instantons on 7-dimensional manifold of G(2) holonomy,''\\
  Mod.\ Phys.\ Lett.\  A {\bf 14} (1999) 2595
  [arXiv:hep-th/9911184];
  %%CITATION = MPLAE,A14,2595;%%

  H.~Kanno and Y.~Yasui,
  ``Octonionic Yang-Mills instanton on quaternionic line bundle of Spin(7)
  holonomy,''
  J. Geom. Phys.  {\bf 34} (2000) 302
  [arXiv:hep-th/9910003];
  %%CITATION = JGPHE,34,302;%%

M.~Dunajski and M.~Hoegner,
  ``SU(2) solutions to self-duality equations in eight dimensions,''
  arXiv:1109.4537 [hep-th].
  %%CITATION = ARXIV:1109.4537;%%
\end{thebibliography}
\end{document}